\documentclass[pdflatex,sn-mathphys-num]{sn-jnl}

\usepackage{graphicx}
\usepackage{multirow}
\usepackage{amsmath,amssymb,amsfonts}
\usepackage{amsthm}
\usepackage[title]{appendix}
\usepackage{xcolor}
\usepackage{textcomp}
\usepackage{booktabs}
\usepackage{threeparttable}
\usepackage{subcaption}
\usepackage{float}
\usepackage{makecell}
\usepackage{lineno}

\begin{document}

\title[Test of the JUNO 20-inch PMTs during Installation]{Test of the JUNO 20-inch PMTs during Installation}

\author[1,2]{\fnm{Haojie} \sur{Dong}}
\author[1]{\fnm{Zhaoyuan} \sur{Peng}}
\author*[1]{\fnm{Zhonghua} \sur{Qin}}\email{qinzh@ihep.ac.cn}
\author[1]{\fnm{Wan} \sur{Xie}}
\author[1]{\fnm{Haoqi} \sur{Lu}}
\author[1]{\fnm{Jun} \sur{Hu}}
\author[1]{\fnm{Lei} \sur{Fan}}
\author[1]{\fnm{Xiaoshan} \sur{Jiang}}
\author[1]{\fnm{Chao} \sur{Chen}}
\author[1]{\fnm{Xiaolu} \sur{Ji}}
\author[1]{\fnm{Fei} \sur{Li}}
\author[1]{\fnm{Shenghui} \sur{Liu}}
\author[1]{\fnm{Xiaochuan} \sur{Xie}}
\author[1]{\fnm{Mei} \sur{Ye}}
\author[1]{\fnm{Hongzhao} \sur{Yu}}
\author[1]{\fnm{Zeyuan} \sur{Yu}}

\affil[1]{\orgdiv{Institute of High Energy Physics}, \orgname{Chinese Academy of Sciences}, \orgaddress{\city{Beijing}, \postcode{100049}, \country{China}}}
\affil[2]{\orgdiv{University of Chinese Academy of Sciences}, \orgname{Chinese Academy of Sciences}, \orgaddress{\city{Beijing}, \postcode{100049}, \country{China}}}

\abstract{Photomultiplier tubes (PMTs) are widely used in neutrino experiments. As a new-generation neutrino observatory, JUNO requires an excellent energy resolution of 3\% at 1~MeV. This will be realized with a 20~kton liquid scintillator detector instrumented with more than 20000 20-inch PMTs and 25600 3-inch PMTs. These PMTs were successfully installed in JUNO from October 2022 to December 2024. During the installation, seven test campaigns were performed to validate the PMT functionality, including measurements dark count rate, gain, waveform, and charge spectrum. In this paper, we present the implementation of these tests and the corresponding results for the 20-inch PMTs throughout the installation process.}


\maketitle

\section{Introduction}\label{sec1}

\begin{figure}[H]
\centering
\includegraphics[width=1\textwidth]{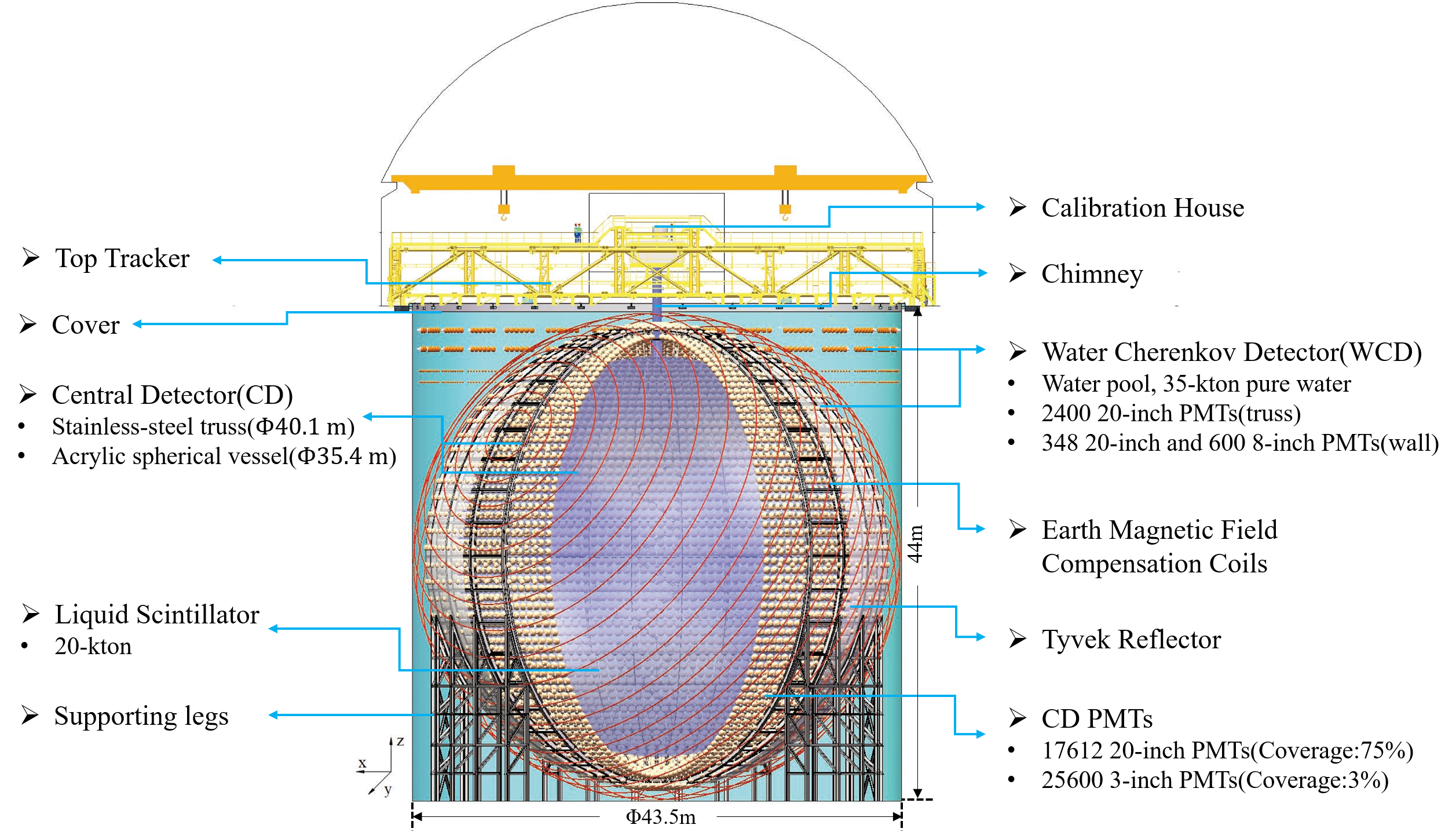}
\caption{Schematic View of the JUNO detector}\label{fig:1}
\end{figure}

The JUNO (Jiangmen Underground Neutrino Observatory)~\cite{JUNO_detector}, which aims to determine the neutrino mass ordering, perform precise measurements of oscillation parameters, and conduct other neutrino studies, is located approximately 700 meters underground in Jiangmen City, Guangdong Province, China. As the world's largest liquid scintillator neutrino experiment to date, JUNO has constructed its giant detector over the past ten years and officially started data taking in August 2025. As illustrated in Figure~\ref{fig:1}, the JUNO detector is composed of several subsystems, mainly a Central Detector and a VETO detector.

The Central Detector (CD)~\cite{JUNO_CD} consists of a spherical acrylic vessel with a diameter of 35.4~m, which holds 20~kton of the liquid scintillator (LS). The acrylic vessel is supported by a stainless-steel truss (SS truss) with a diameter of 40.1~m. A total of 17612 20-inch PMTs and 25,600 3-inch PMTs are designed to be deployed on the inner surface of the SS truss, facing inwards to the CD center to detect signals produced by neutrinos interactions with the LS.

The VETO system, comprising a Water Cherenkov Detector (WCD)~\cite{JUNO_WCD} and a Top Tracker~\cite{JUNO_TT}, is designed primarily for cosmic-ray muon tagging. The WCD also helps suppress background from environmental radioactivity. The WCD is a cylindrical water pool with a diameter of 44~m and a height of 43.5~m, filled with 35~kton high-purity water. It is designed to instrument 2400 20-inch PMTs on the outer surface of the SS truss, 348 20-inch PMTs and 600 8-inch PMTs on the water pool wall. In addition, a black plastic cover at the top of water pool provides optical sealing for all PMTs.

In total, 20360 20-inch PMTs are planned for JUNO, including 17612 CD PMTs and 2748 VETO PMTs. Of these, 15360 are MCP PMTs from NNVT~\cite{NNVT_PMT}, and 5000 are dynode PMTs from Hamamatsu~\cite{Hamamatsu_R12860}. The installation of these 20-inch PMTs spanned more than two years, from October 2022 to December 2024, with 20343 PMTs successfully installed in total. Seven rounds of PMT test campaigns were conducted throughout the installation process, and are presented in this paper. In the following, Section~\ref{sec2} provides a brief overview of the 20-inch PMT installation, followed by a detailed description of the PMT tests in Section~\ref{sec3}.

\section{Installation of the 20-inch PMTs}\label{sec2}

\begin{figure}[H]
\centering
\begin{subfigure}[t]{0.48\textwidth}
    \centering
    \includegraphics[height=5cm,keepaspectratio]{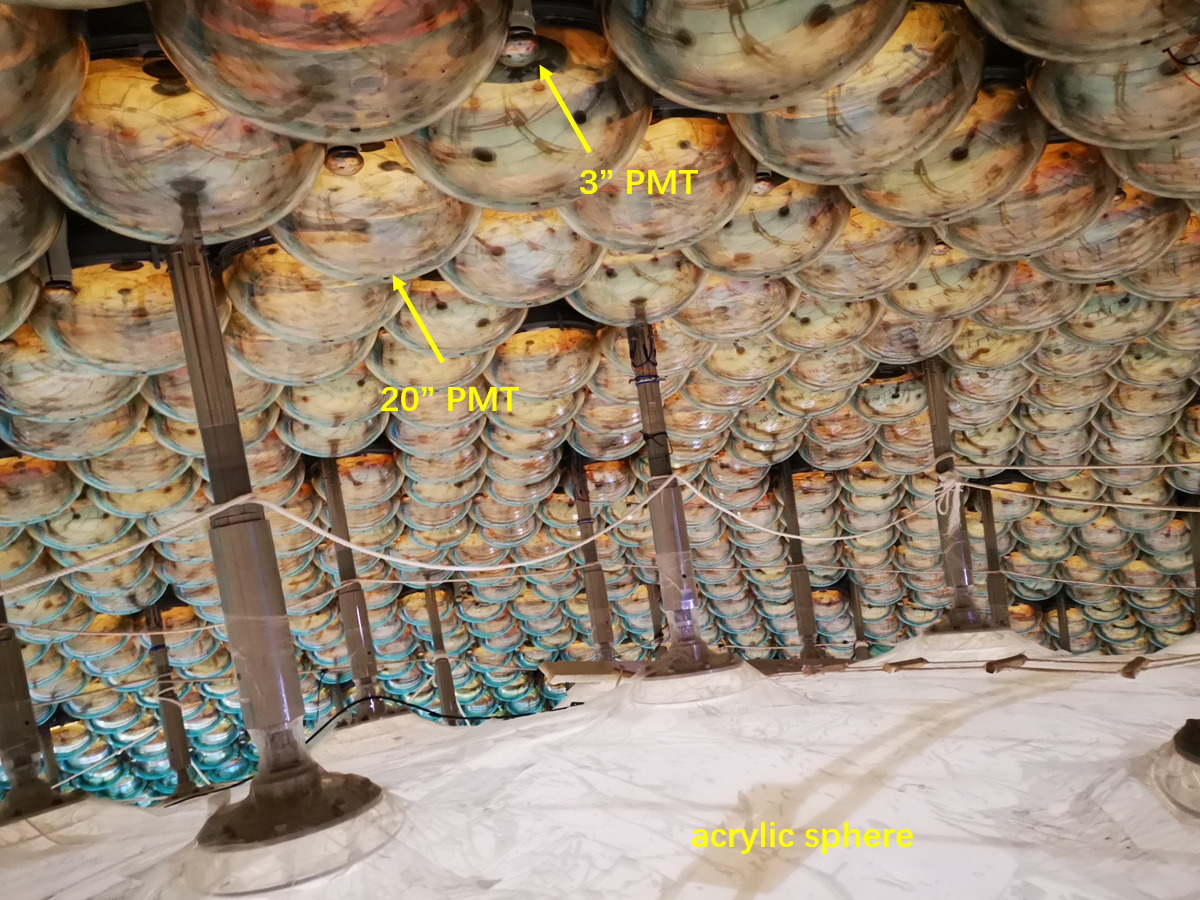}
    \caption{}
\end{subfigure}
\hfill
\begin{subfigure}[t]{0.48\textwidth}
    \centering
    \includegraphics[height=5cm,keepaspectratio]{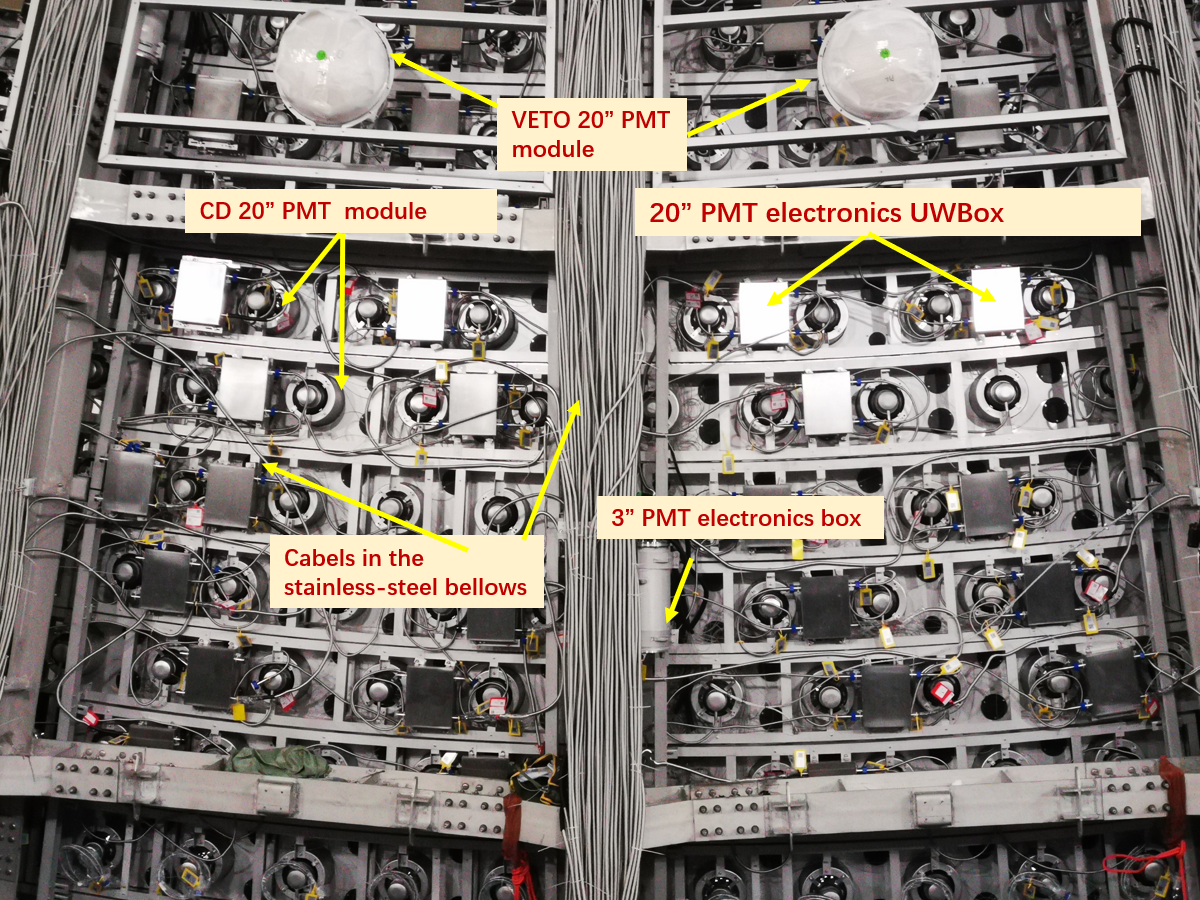}
    \caption{}
\end{subfigure}
\caption{Installed PMTs and associated equipment. (a) PMTs on the inner surface of the SS truss; (b) PMT modules, electronics and cables on the outer surface of the SS truss.}
\label{fig:installed}
\end{figure}

To meet the JUNO's stringent energy resolution requirement (3\% at 1~MeV), the CD 20-inch PMTs must achieve an optical coverage of 75\%. This requires an extremely compact PMT arrangement with an average gap of only 3~mm between adjacent PMTs. To achieve such high density and high precision, a modularized installation approach was adopted for both CD and VETO PMTs. In the JUNO surface assembly building, PMTs were firstly assembled into modules, each hosting 1 to 8 PMTs depending on the module size, before being transported to the underground experiment hall. Installation of the PMT modules proceeded in a top-to-bottom sequence on the SS truss, and several zones were designated to accommodate different installation methods and platforms. In parallel with 20-inch PMT installation, the 20-inch PMT electronics, cables, 3-inch PMTs and their electronics, and other equipment were also installed. Figure~\ref{fig:installed}(a) shows the CD 20-inch PMTs (including 3-inch PMTs) on the inner surface of the SS truss, between the acrylic vessel and the SS truss. Figure~\ref{fig:installed}(b) shows 20-inch PMT modules (including CD and VETO modules), 20-inch PMT under water electronics (housed in a stainless steel box called the under water box, UWBox), cables (encapsulated in a stainless steel bellows), and 3-inch PMT under water electronics (also inside a stainless-steel box) on the outer surface of the SS truss. A more detailed description of the PMT installation is currently in preparation as a dedicated publication.

In summary, from October 23, 2022 to December 18, 2024, a total of 20,343 20-inch PMTs were successfully installed, including 17596 CD PMTs and 2747 VETO PMTs. Relative to the planned 20360 PMTs, 17 PMTs could not be installed due to spatial conflicts either between PMTs or between PMTs and other components. The impact of these uninstalled PMTs is minor: the final optical coverage of CD 20-inch PMTs reached 75.14\%, which satisfies the JUNO design requirement of 75\%.

\section{Test of the 20-inch PMTs during Installation}\label{sec3}

\subsection{Motivations of the test}

Firstly, due to the highly compact layout of PMTs and associated equipment, it is very difficult, or even impossible, to replace the earlier-installed PMTs once many subsequent layers have been installed. Therefore, regular tests after the installation of several PMT layers are essential to promptly verify the PMT functionality and identify problems from the installation, so that the problems can be solved in a timely manner.

Secondly, since a lot of associated equipment is installed at the same time, these tests can serve as joint tests across different systems, including PMTs, electronics, trigger, DAQ (data acquisition system) and DCS (detector control system), to investigate and solve potential issues on a broader scale.

Lastly, it's also crucial to have initial operational practice on such a large number of PMTs and associated equipment at an early stage, in order to gain experience for the subsequent commissioning and official operation.

\subsection{Test protocol and setup}

\begin{figure}[H]
\centering
\includegraphics[width=1\textwidth]{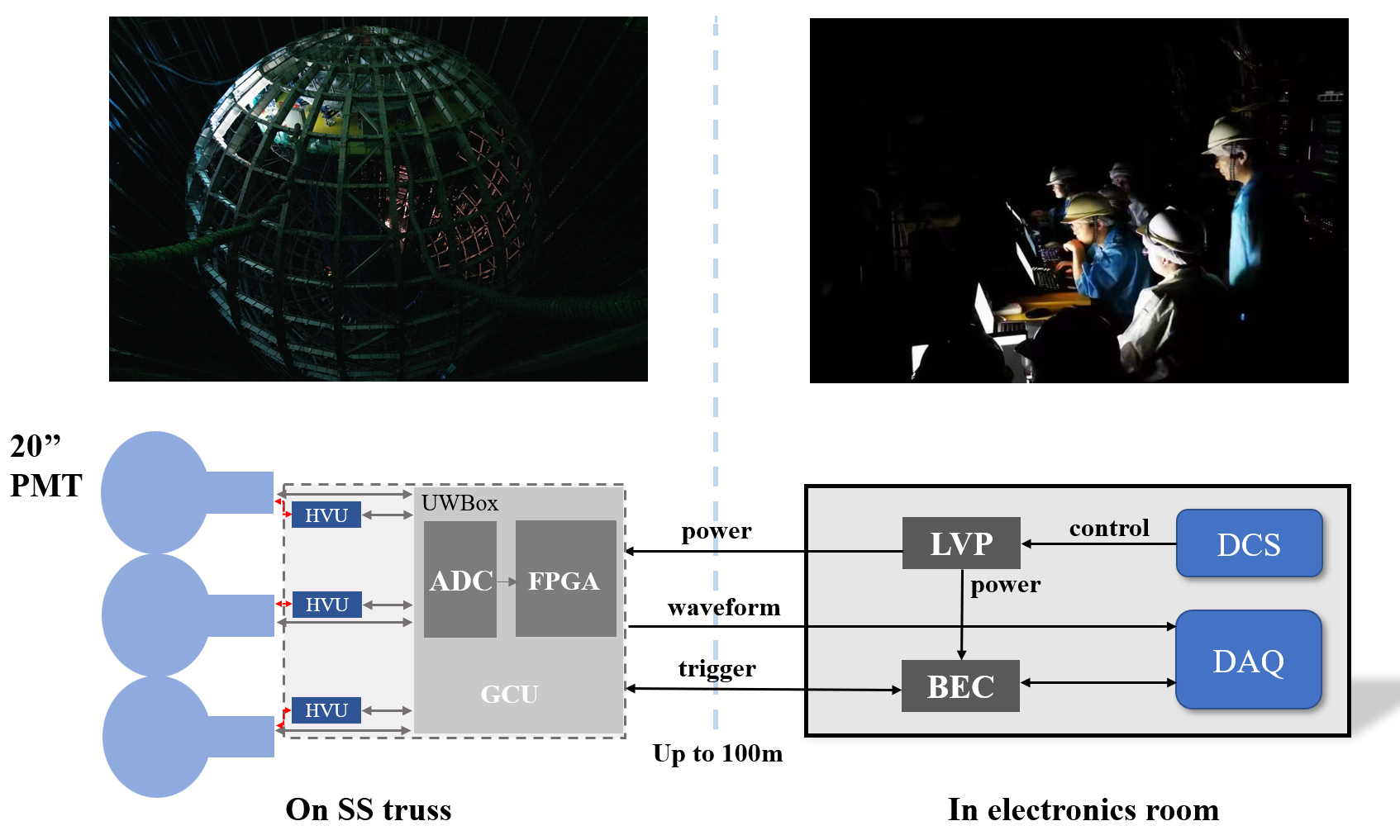}
\caption{Photos and schematic diagrams of the test setup. Top left: a photo of the experimental hall and SS truss during the light-off procedure; top right: a photo of test operations in the electronics room; bottom left: a schematic of the PMTs and underwater electronics, with one UWBox connecting to three PMTs; bottom right: a schematic of the remaining electronics components, DAQ, and DCS systems.}
\label{fig:test_system}
\end{figure}

Photos and schematic diagrams of the test setup are shown in Figure~\ref{fig:test_system}: on the SS truss, where the PMTs and electronics UWBoxes are located, PMT high voltages are supplied by HVUs (High Voltage Units). For the signal readout, the PMT dark noise waveforms are continuously digitized by 1~GS/s (samples per second) Flash ADCs in 1000~ns window. Concurrently, the DCR is dynamically computed by a self-trigger logic running in the embedded FPGA of the GCU (Global Control Unit): any signal exceeding a dedicated threshold---15~ADC counts based on earlier threshold scanning~\cite{Peng_2025}---is registered as a valid dark count. The FPGA also buffers the corresponding waveform data. In the electronics room, the LVP (Low Voltage Power supply) provides power to HVUs under DCS control, BEC (Back-End Card) sends trigger to GCU, and data are acquired by the DAQ system. A detailed description of the electronics, DAQ and DCS can be found in~\cite{Petitjean_2022,COPPI2023168255,JUNO_DAQ,JUNO_DCS}.

To conduct the PMT test, a completely dark environment is required. To this end, all visible light sources surrounding the PMTs---including the SS truss, experimental hall, nearby rooms and tunnels---must be turned off. To ensure no visible light spots exist, a dedicated 20-inch PMT placed in the experimental hall was turned on to serve as a monitor. 
If the dark count rate (DCR) from this monitoring PMT was in a reasonable range, the PMTs to be tested on the SS truss were then powered on. The PMT high voltages were increased step by step from 0~V, to 800~V, 1200~V, 1600~V, and eventually to their design values. At each intermediate step, the high voltages were kept for several minutes and the PMT DCRs were measured. If the DCRs were too high ($>$ 600~kHz for most of the PMTs), the test was paused and further inspections on light sources and other causes were performed. This stepwise method is designed to mitigate potential damage to the PMTs caused by unexpected factors, such as residual light in the surrounding area, PMT instability due to initial power-on, or problems from installation. In the final step, the high voltages were adjusted to the values corresponding to a gain of $1.0 \times 10^{7}$ based on the previous test results in the Pan-Asia warehouse~\cite{PMT_mass_test,Wonsak_2021}. This step was held relatively longer duration, up to two or three hours depending on the number of PMTs and list of items to be tested.

In addition, owing to the tight installation schedule, tests were typically scheduled late at night to allow installation work to resume the following day. Furthermore, because of the large surrounding area and complex installation conditions, several hours were taken to achieve light blocking, and a second inspection of light sources was usually needed during the test. For these reasons, frequent testing was not feasible. In total, seven test campaigns were carried out over the installation period. Detailed information for each test, including the date, corresponding PMT layers, number of PMTs tested, and cumulative totals, is listed in Table~\ref{tab:4}. Altogether, 19,843 PMTs were tested in these campaigns. The remaining 500 units (20,343 $-$ 19,843) were installed at the very final stage and measured later during the water-filling phase; they are therefore not included in this table.

\begin{table}[htbp]
    \centering
        \caption{Detailed information for all tests during installation}
    \label{tab:4}
    \begin{tabular}{lcccccc}
    \toprule
    \multirow{2}{*}{\textbf{}} & \multirow{2}{*}{\textbf{Test time}} & \multicolumn{2}{c}{\textbf{CD PMT}} & \multicolumn{2}{c}{\textbf{VETO PMT}} & \textbf{Total} \\
    \cmidrule(lr){3-4} \cmidrule(lr){5-6}
    & & Layer ID & PMT Numbers & Layer ID & PMT Numbers & Numbers \\
    \midrule
    1st & 09/12/2022 & N60-N57 & 57 & -- & 0 & 57 \\
    2nd & 17/04/2023 & N60-N48 & 682 & N11 & 15 & 697 \\
    3rd & 05/09/2023 & N60-N20 & 4826 & N11-N05 & 465 & 5291 \\
    4th & 26/05/2024 & N60-N13 & 6203 & N11-N03 & 660 & 6863 \\
    5th & 10/08/2024 & N12-S06 & 3954 & N02-N01 & 162 & 4116 \\
    6th & 28/09/2024 & S07-S25 & 3957 & S01-S04 & 906 & 4863 \\
    7th & 17/11/2024 & S26-S60 & 3714 & S05-S08 & 575 & 4289 \\
    \midrule
    \multicolumn{2}{l}{In total:} & & 17554 & & 2289 & 19843 \\
    \bottomrule
    \end{tabular}
    \vspace{1ex}
\par\begingroup\footnotesize N refers to the north hemisphere of the SS truss, and S refers to the south hemisphere. For CD, PMTs are arranged from N62 to N1 on the north hemisphere, and from S62 to S1 on the south hemisphere. While for VETO, PMTs are arranged from N12 to N1 on the north hemisphere and from S12 to S1 on the south hemisphere. The sum of the PMT numbers in each test is slightly larger than the total number because some PMTs were tested multiple times.
    \endgroup
    \end{table}

\subsection{Test Results}\label{sec:results}

\subsubsection{Prompt results}

\begin{figure}[H]
\centering
\includegraphics[width=1\textwidth]{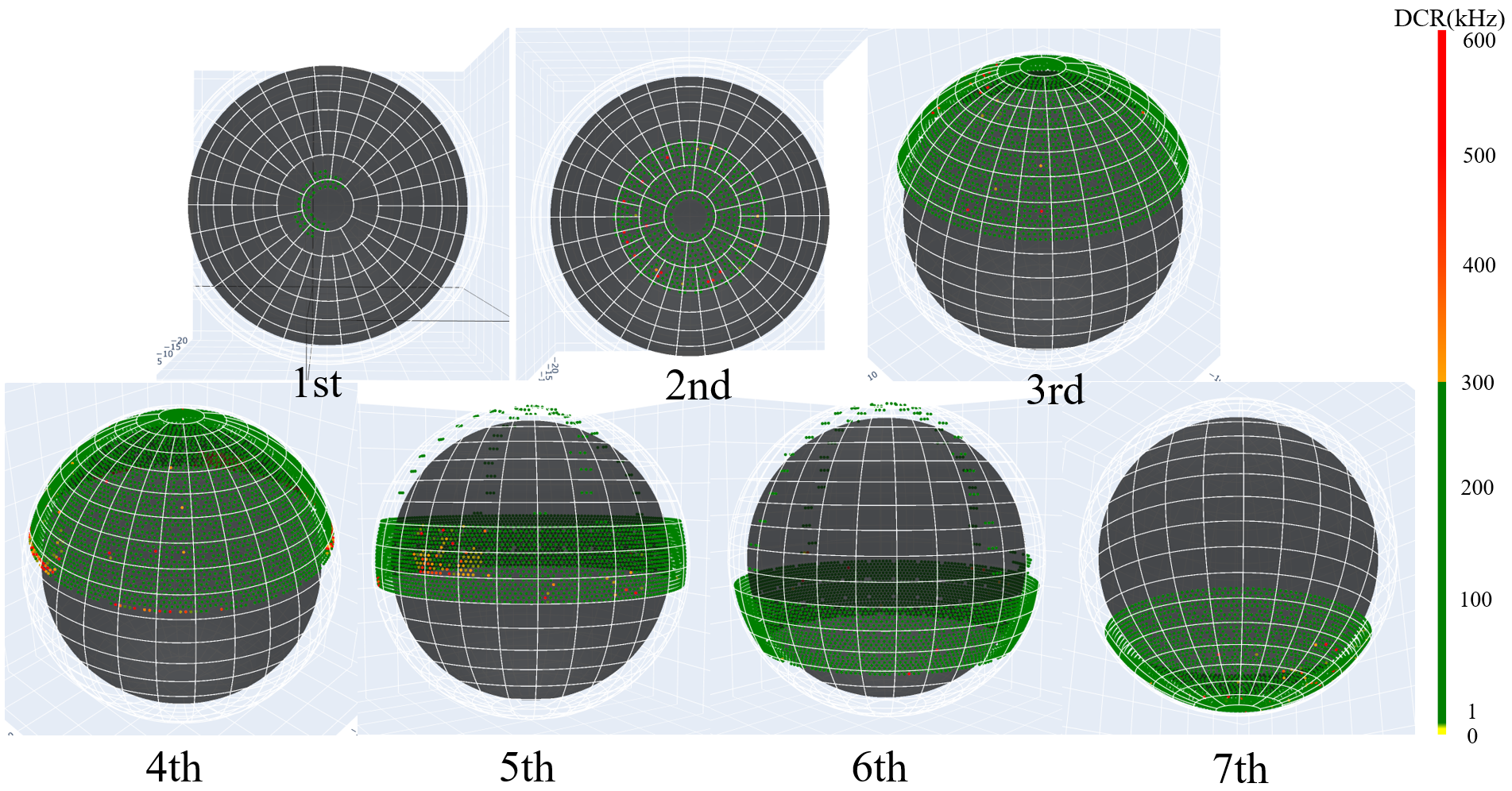}
\vspace{0.5cm}
\caption{3D PMT maps for all test campaigns}
\label{fig:pmtmap}
\end{figure}

As mentioned above, the PMT DCRs are calculated by the FPGA embedded in GCU, so the DCR results can be provided in real time. This allows the PMT functionality to be verified promptly, even during the test process. If any issues are found, they can be addressed immediately. A three-dimensional map of the PMTs for each test is shown in Figure~\ref{fig:pmtmap}, where each dot represents a PMT installed on the SS truss, and the color indicates its DCR. Green dots denote PMTs with DCR $<$ 300~kHz, while other colors represent PMTs with DCR $>$ 300~kHz. It can be seen that the vast majority of PMTs in all tests exhibited DCRs below 300~kHz. A small fraction of ``hot'' PMTs appeared in the 4th and 5th tests: those in the 4th test were distributed around the equator, attributed to residual light on the equatorial platform, while those in the 5th test were concentrated, caused by a light source on the SS truss. There are also some isolated PMTs with very high DCRs, which occurs because certain PMTs are more sensitive to light exposure and require longer cool-down time. From these maps, it can be concluded that the PMTs functioned properly after installation, and no major problems were identified during the installation process. However, it should be noted that at this stage, a DCR of 300~kHz or even higher is reasonable and acceptable, owing to incomplete light blocking and short cool-down time.

\begin{figure}[H]
\centering
\begin{subfigure}[b]{0.48\textwidth}
    \includegraphics[width=\textwidth]{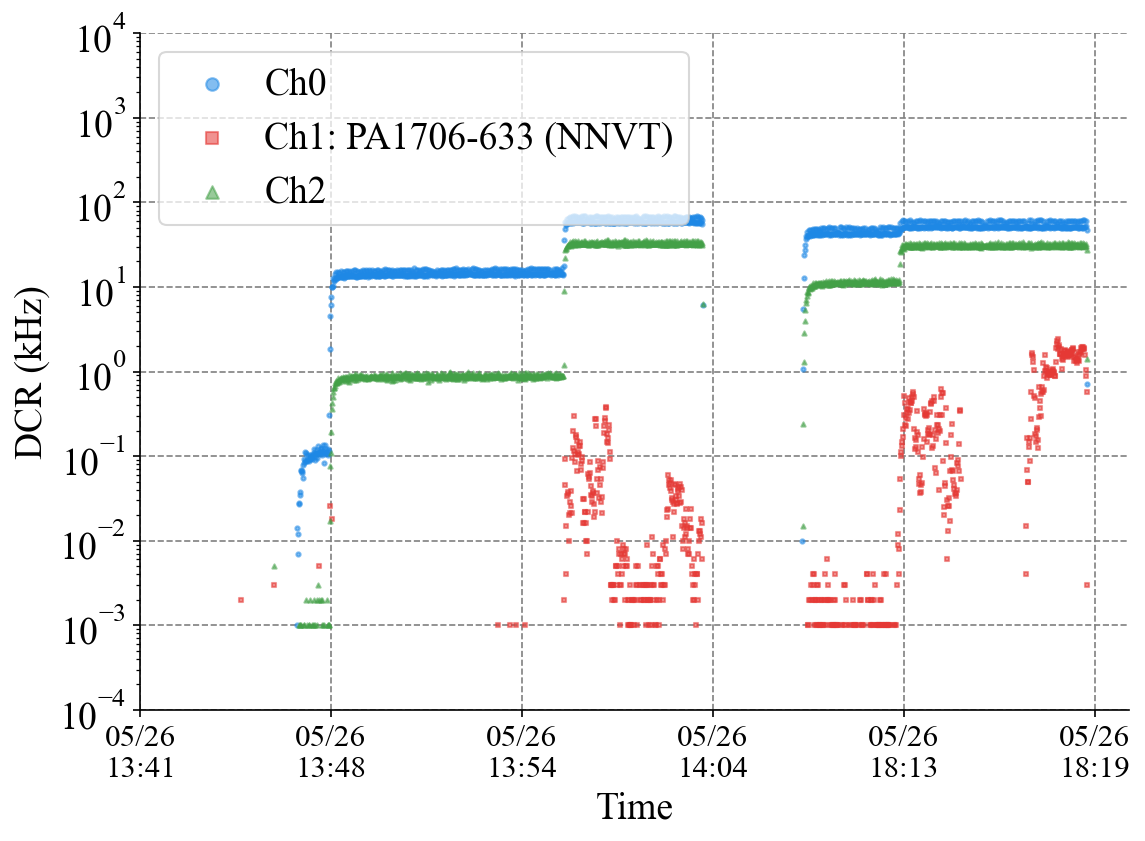}
    \caption{}
\end{subfigure}
\hfill
\begin{subfigure}[b]{0.48\textwidth}
    \includegraphics[width=\textwidth]{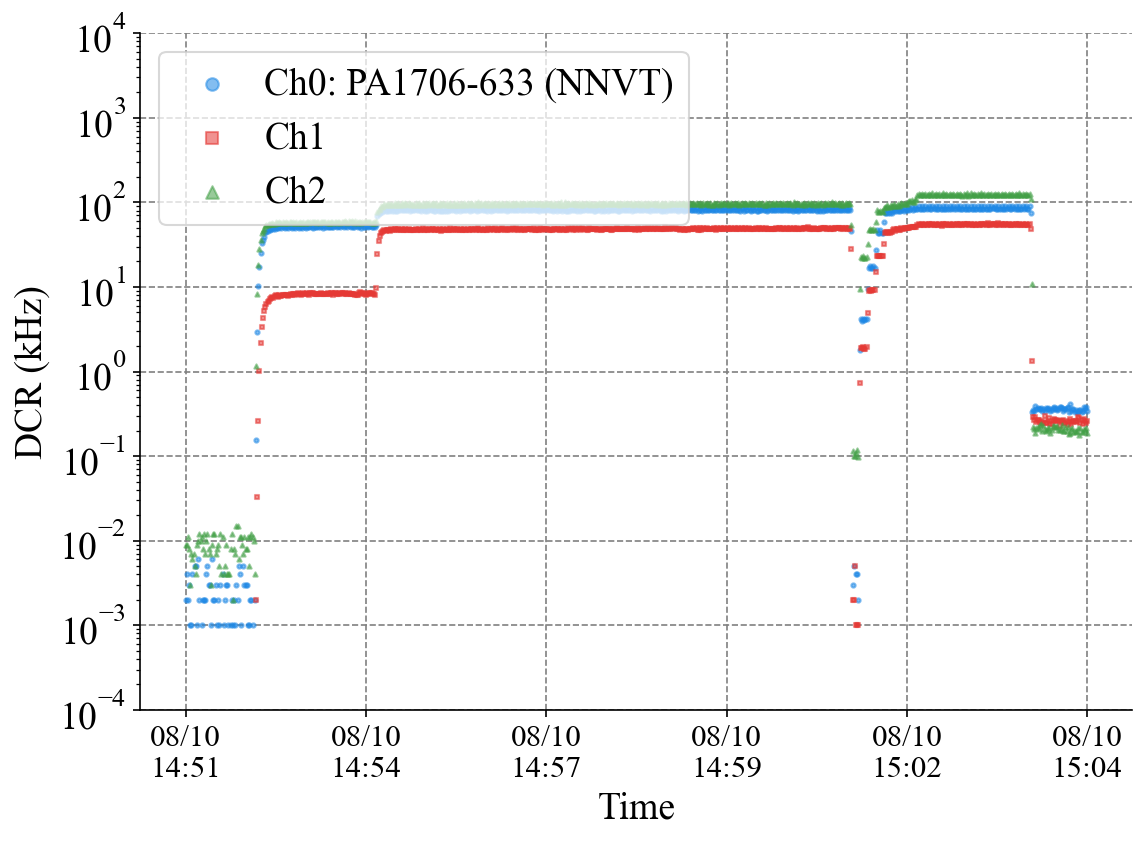}
    \caption{}
\end{subfigure}
\caption{PMT response recovered after channel swap. (a) The PMT of PA1706-633 connected to channel 1 of the UWBox shows no response with an extremely low DCR; (b) after swapping to channel 0, the DCR returns to normal.}
\label{fig:recovery}
\end{figure}

After a detailed inspection of these maps, four PMTs were found to show no response after high voltage was applied, i.e. with DCR of 0 or an extremely low value. To diagnose the issue, the electronics readout channels of these PMTs were swapped with those of neighboring channels. Following the channel swap, two of the four PMTs recovered their response, while the other two did not. This issue was attributed to poor contact between the PMTs and the readout electronics during installation. Given the extreme difficulty of accessing these non-recovered PMTs and their small number (only two), it was decided not to replace them. Figure~\ref{fig:recovery} shows an example of a recovered PMT: after swapping of channels within the same UWBox, the previously unresponsive PMT recovered its response and exhibited a normal DCR.

In addition to the prompt results presented above, a subsequent analysis of the basic PMT performance parameters is provided in the following sections, covering the PMT gain, DCR, waveform and charge spectrum. These parameters were analysed from the dark noise data acquired during the installation tests. For other parameters, such as the photon detection efficiency, transit time spread, etc., require a more complicated experimental setup and thus could not be measured during the installation stage. Instead, these quantities were measured during the JUNO commissioning phase, after the full deployment of the detector calibration system.

\subsubsection{PMT Gain}
\label{sec:gain}

\begin{figure}[H]
\centering
\includegraphics[width=0.56\textwidth]{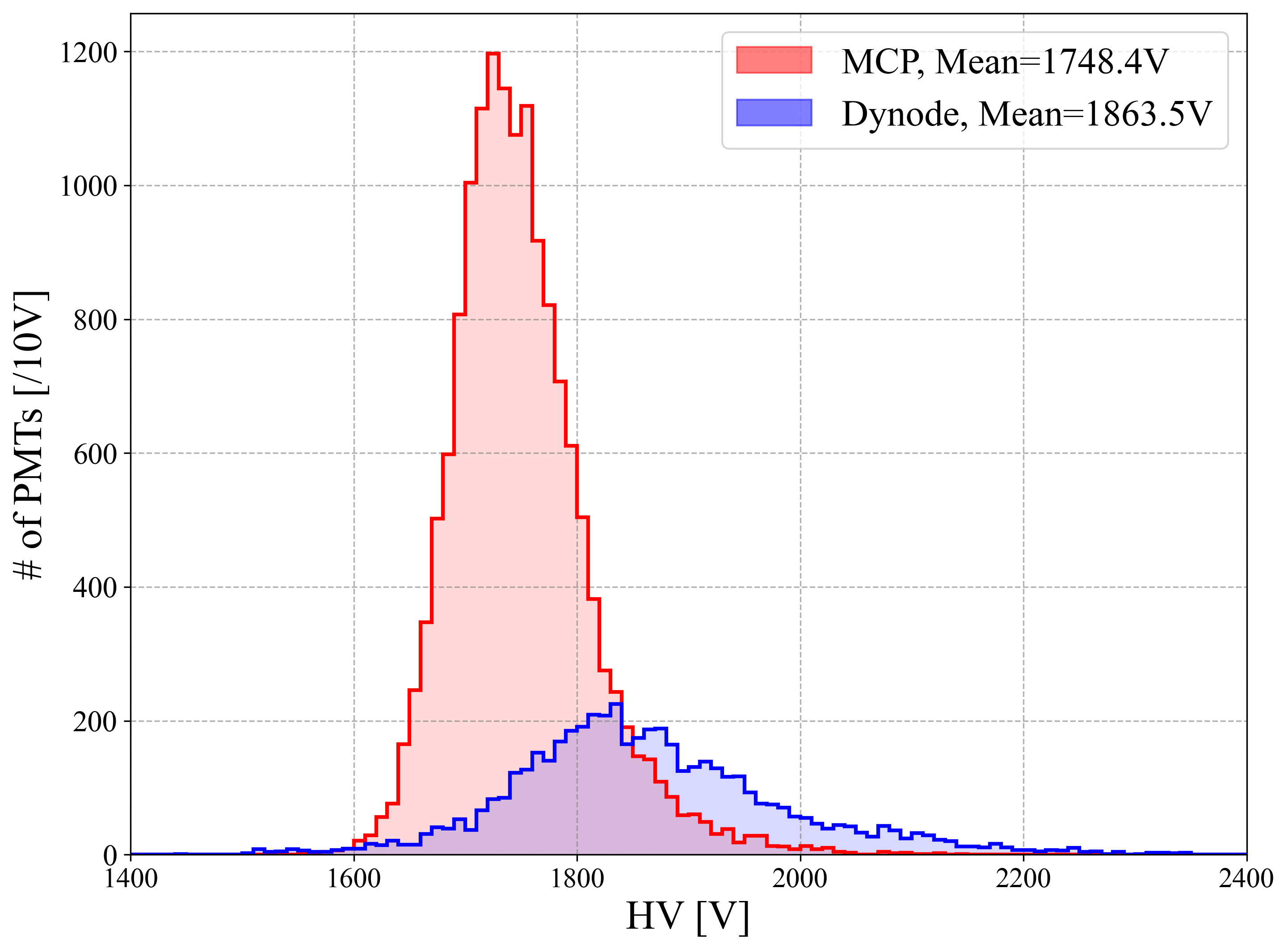}
\caption{High voltage distribution for MCP and dynode PMTs}
\label{fig:12}
\end{figure}

\begin{figure}[H]
\centering
\begin{subfigure}[t]{0.48\textwidth}
    \centering
    \includegraphics[width=\textwidth]{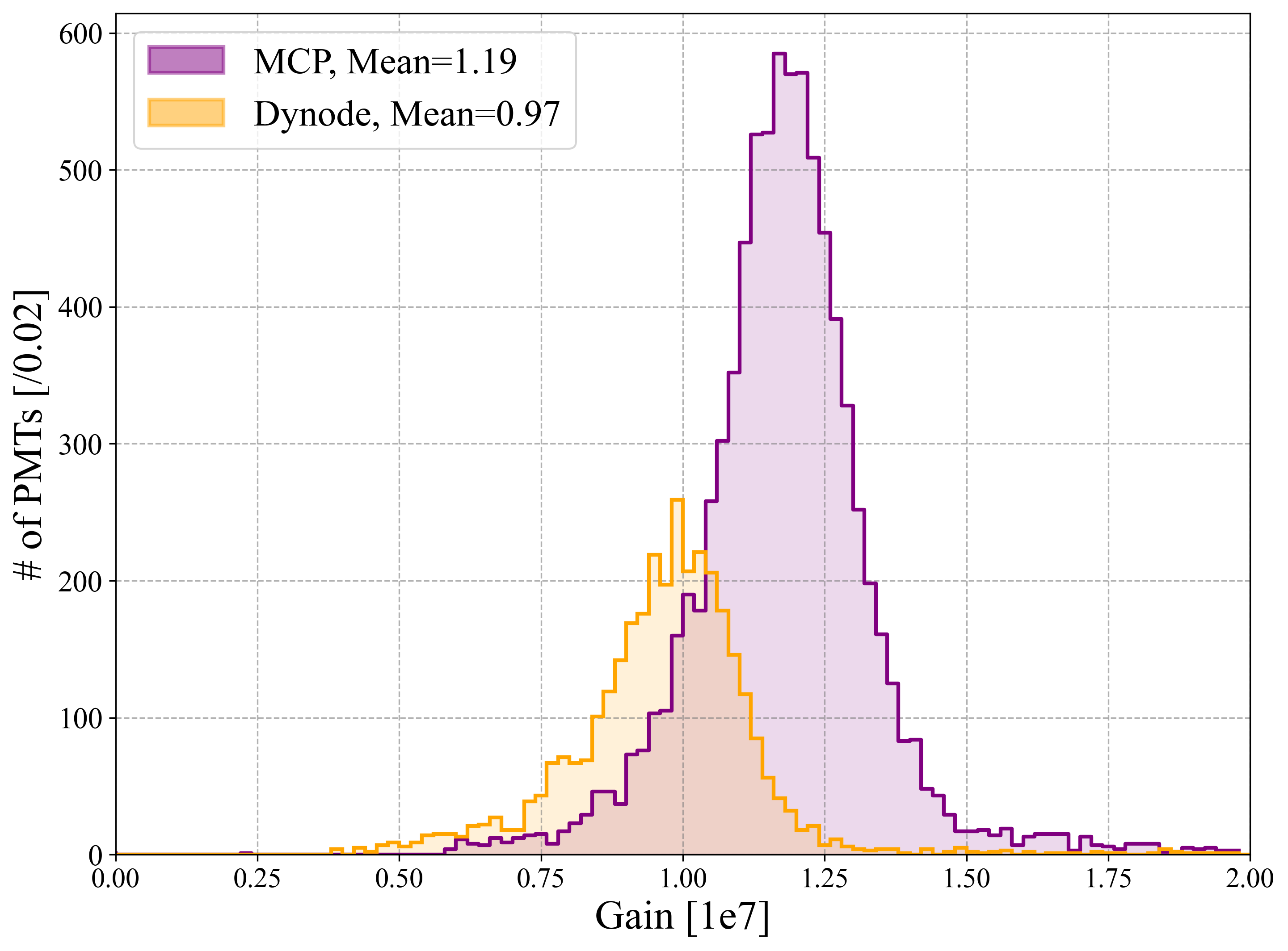}
    \caption{}
    \label{fig:13}
\end{subfigure}
\hfill
\begin{subfigure}[t]{0.48\textwidth}
    \centering
    \includegraphics[width=\textwidth]{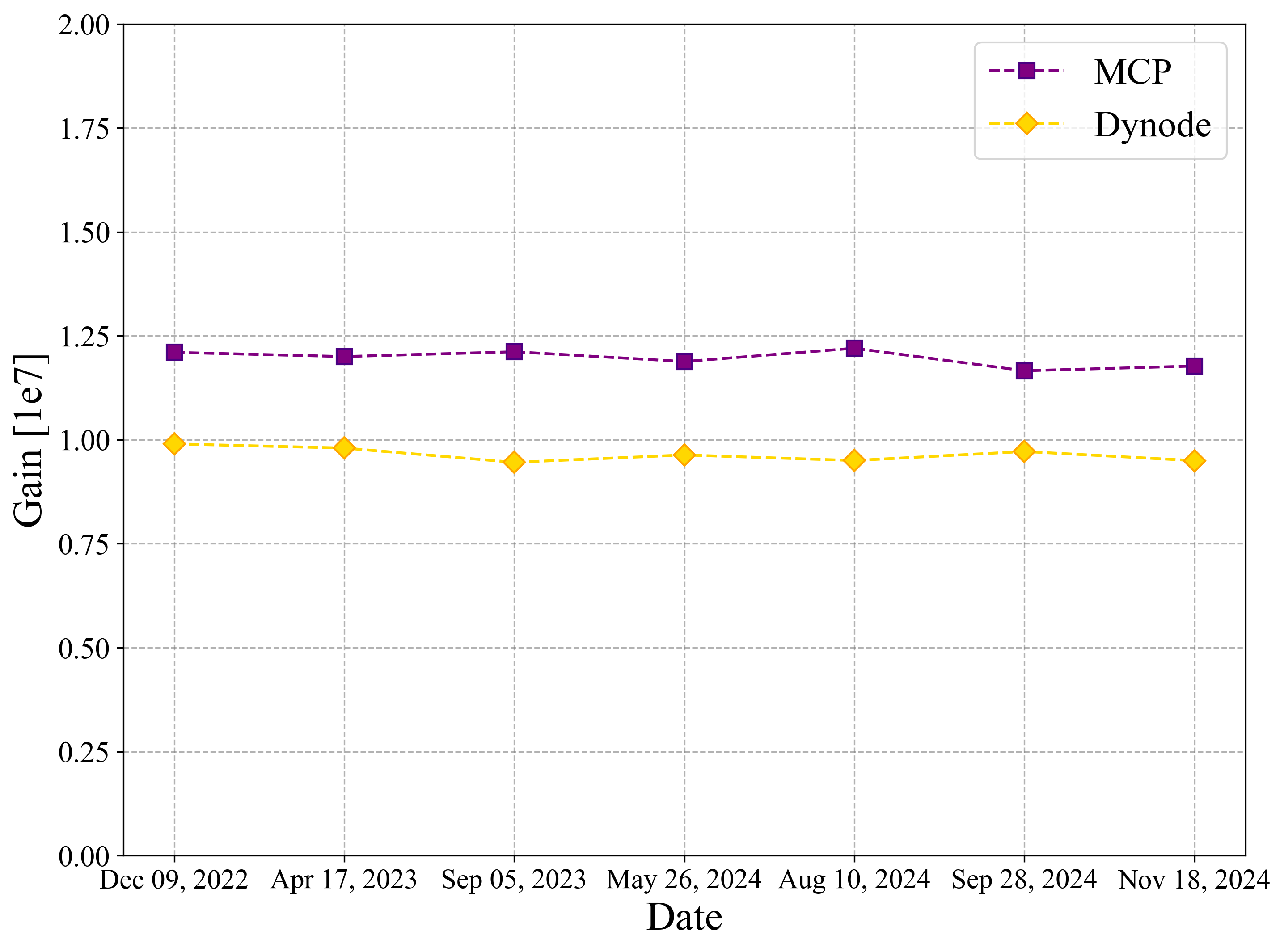}
    \caption{}
    \label{fig:14}
\end{subfigure}
\caption{PMT gain analysis results. (a) Gain distribution of all PMTs across all tests; (b) PMT Gain of each test.}
\label{fig:gain}
\end{figure}

The PMT gain was verified at the same high voltages used in the Pan-Asia test: an average of 1748~V for MCP-PMTs and an average of 1863~V for dynode-PMTs, with the corresponding voltage distribution shown in Figure~\ref{fig:12}. By performing a Gaussian fit to the peak of the dark noise charge spectrum (introduced later in Section~\ref{sec:waveform}), the gain distribution for all PMTs across seven tests and the gain values for each individual test are presented in Figure~\ref{fig:gain}. The average gain is approximately $1.19\times10^{7}$ for MCP PMTs, and approximately $0.97\times10^{7}$ for dynode PMTs. The observed gain deviations from the initial target of $1.0\times10^{7}$ are attributed to the changes in test conditions relative to the Pan-Asia test, such as potting effects, temperature variations, the absence of earth magnetic field, different readout electronics and threshold settings, and potentially altered dark noise charge spectra in the presence of residual ambient light. It should also be noted that the deviation from MCP PMTs is considerably larger than that from dynode PMTs, which may imply that MCP PMTs are more easily affected when test conditions change. However, the gain variation across the tests is small, confirming that all PMTs were subject to the same set of influencing factors and that the deviations are not issues inherent to the PMTs themselves. Nevertheless, the gains will be calibrated to uniform values for both PMT types by adjusting their high voltages during subsequent official operation.

\subsubsection{PMT DCR}

\begin{figure}[H]
\centering
\begin{subfigure}[t]{0.48\textwidth}
    \centering
    \includegraphics[width=\textwidth]{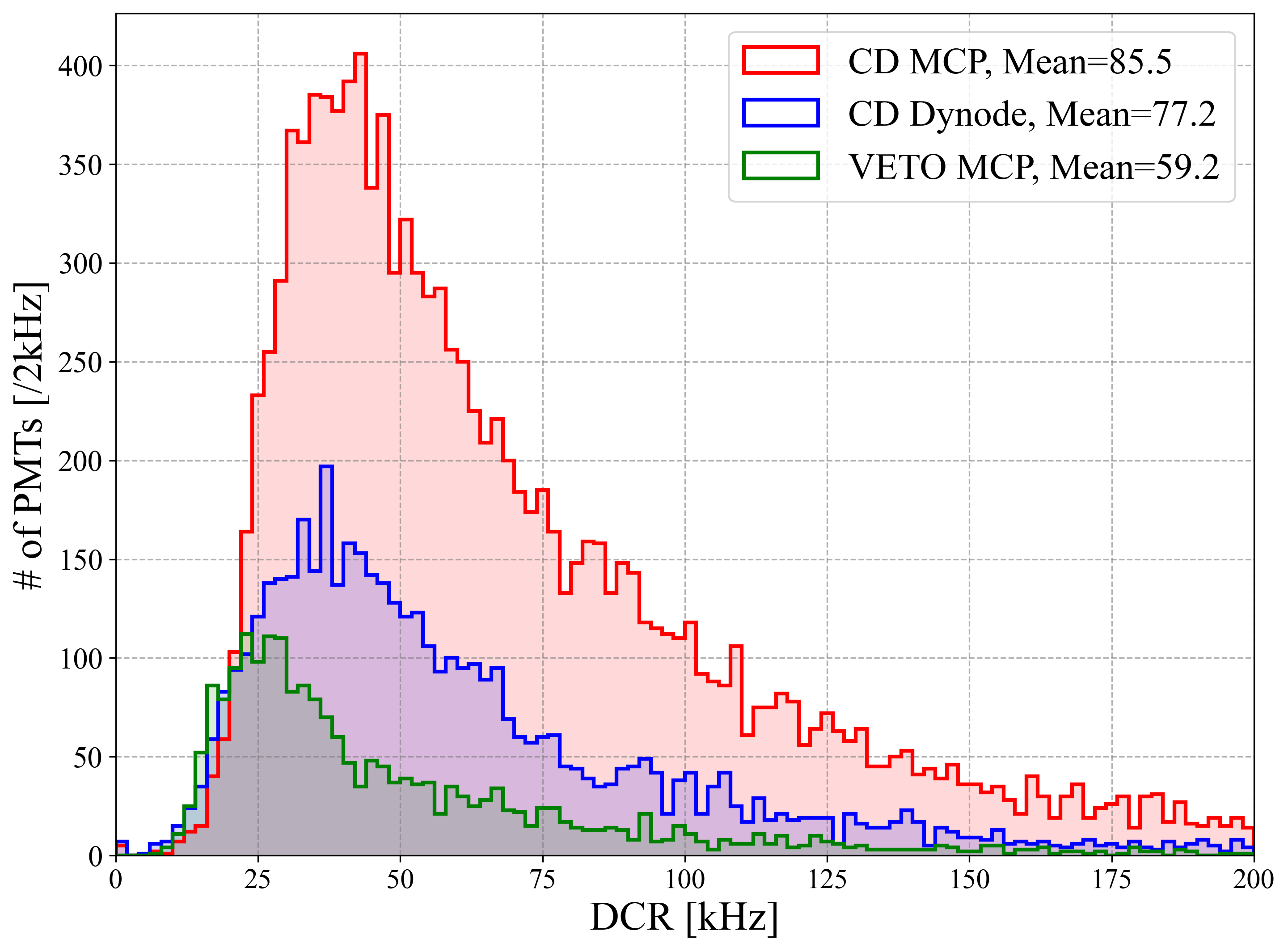}
    \caption{}
\end{subfigure}
\hfill
\begin{subfigure}[t]{0.48\textwidth}
    \centering
    \includegraphics[width=\textwidth]{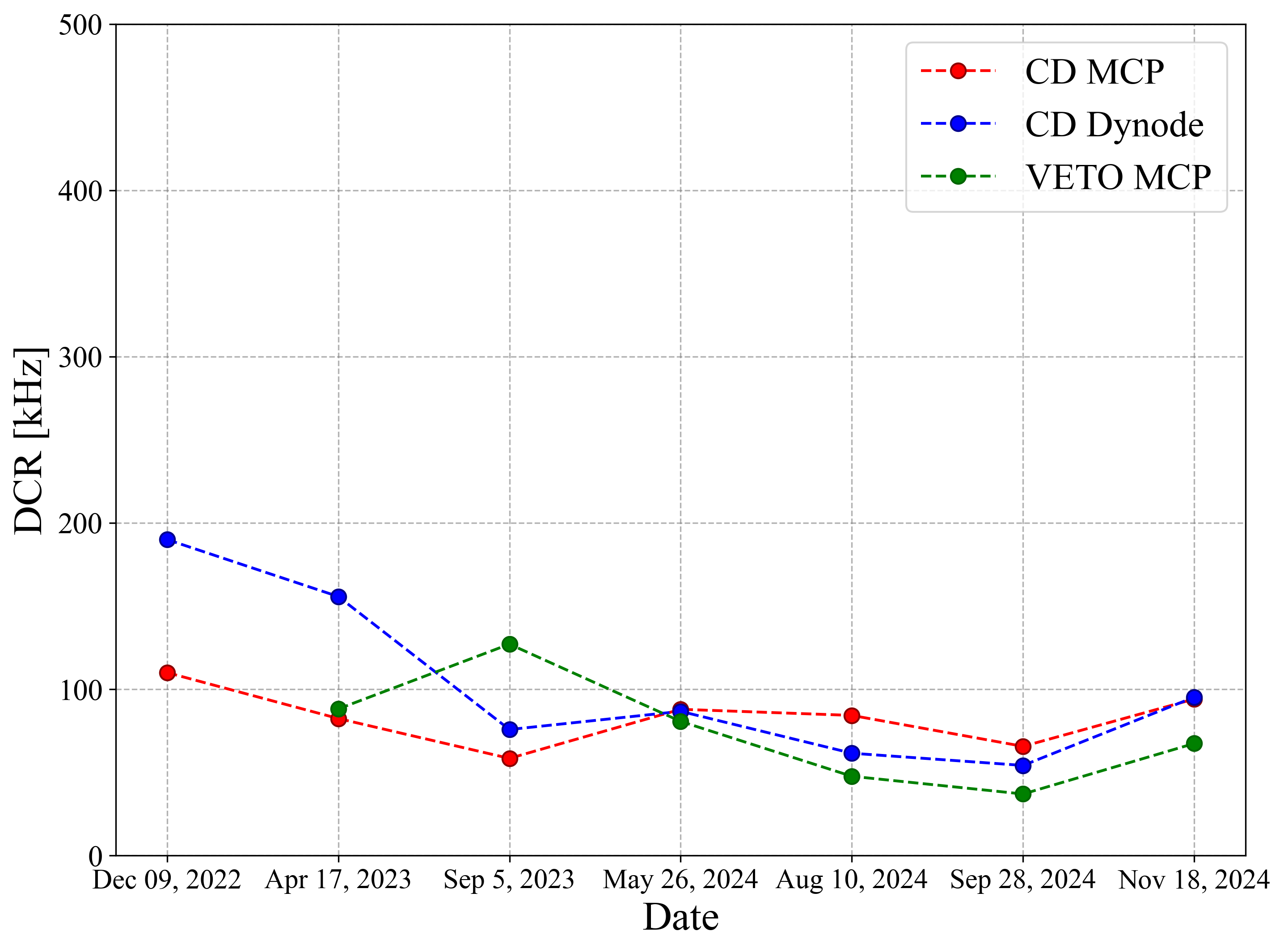}
    \caption{}
\end{subfigure}
\caption{PMT DCR analysis results. (a) The overall distribution of DCR for all PMTs in all tests; (b) DCR values for each individual test.}
\label{fig:dcr}
\end{figure}

In addition to the PMT maps presented previously, Figure~\ref{fig:dcr} shows the overall DCR distributions for all PMTs across all tests, as well as DCR values for each individual test. It can be observed that the mean DCR values are approximately 59~kHz to 85~kHz for different PMT types and positions, with higher DCR values for CD PMTs relative to VETO PMTs due to more residual light coming from the interior where the ambient conditions are more complex; while for different tests, the values can exceed 100~kHz, reaching up to 200~kHz. As noted earlier, the primary reasons for the elevated DCR levels (compared to the $\sim$25~kHz observed in the Pan-Asia test) are attributed to residual ambient light and insufficient cool-down time. Since the residual light intensity, which was difficult to control, varied across each test, the DCR exhibited significant fluctuations. It is expected that the DCR will decrease substantially during the official JUNO operation, once the PMTs have undergone sufficient cooling and the ambient light has been completely blocked by the black plastic cover at the top of water pool. This expectation is confirmed by the results reported in the recently published JUNO detector performance paper~\cite{JUNO_detector_performance}.

\subsubsection{PMT Waveform and Charge Spectra}
\label{sec:waveform}

\begin{figure}[H]
\centering
\begin{subfigure}[t]{0.48\textwidth}
    \centering
    \includegraphics[width=\textwidth]{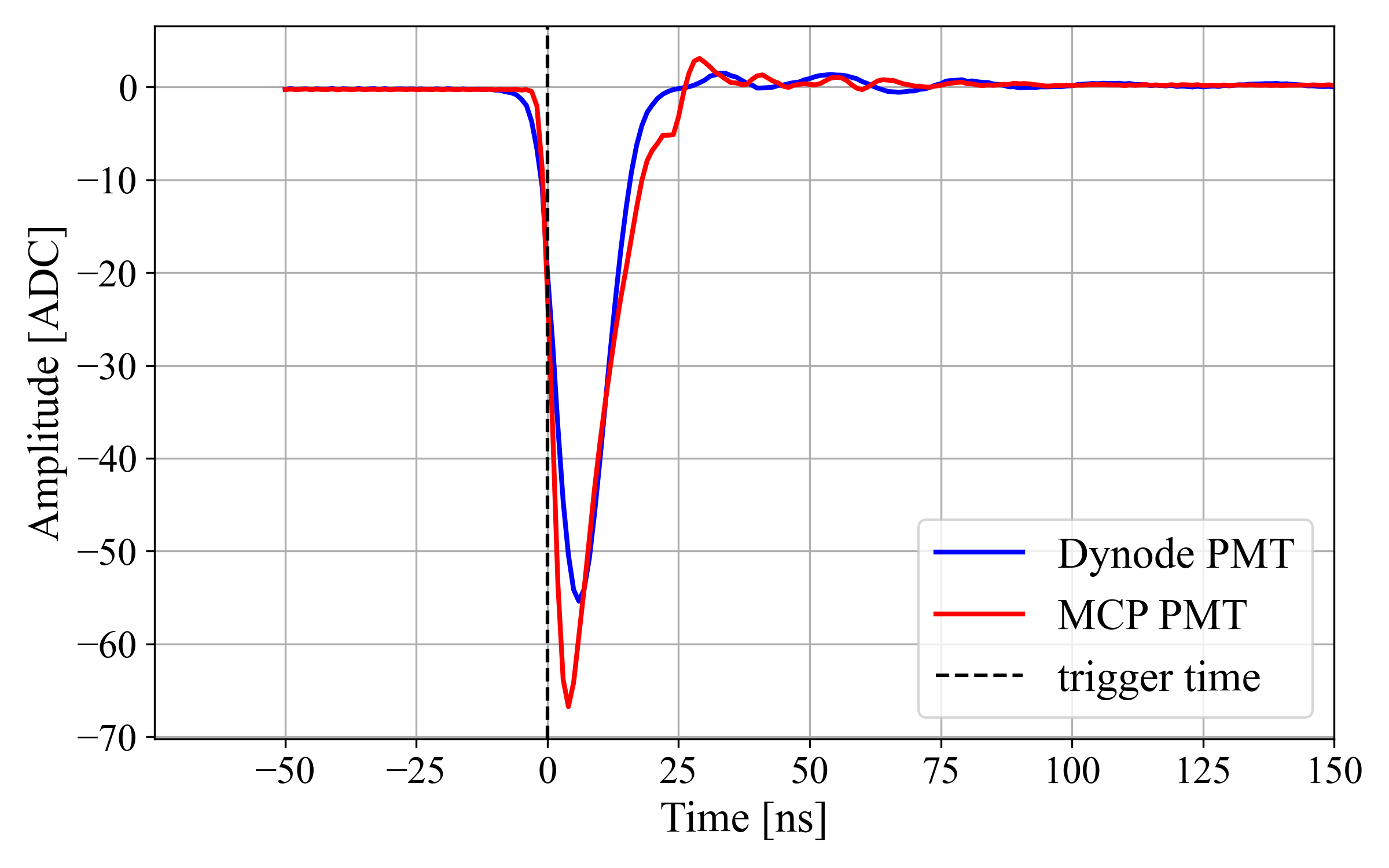}
    \caption{}
\end{subfigure}
\hfill
\begin{subfigure}[t]{0.48\textwidth}
    \centering
    \includegraphics[width=\textwidth]{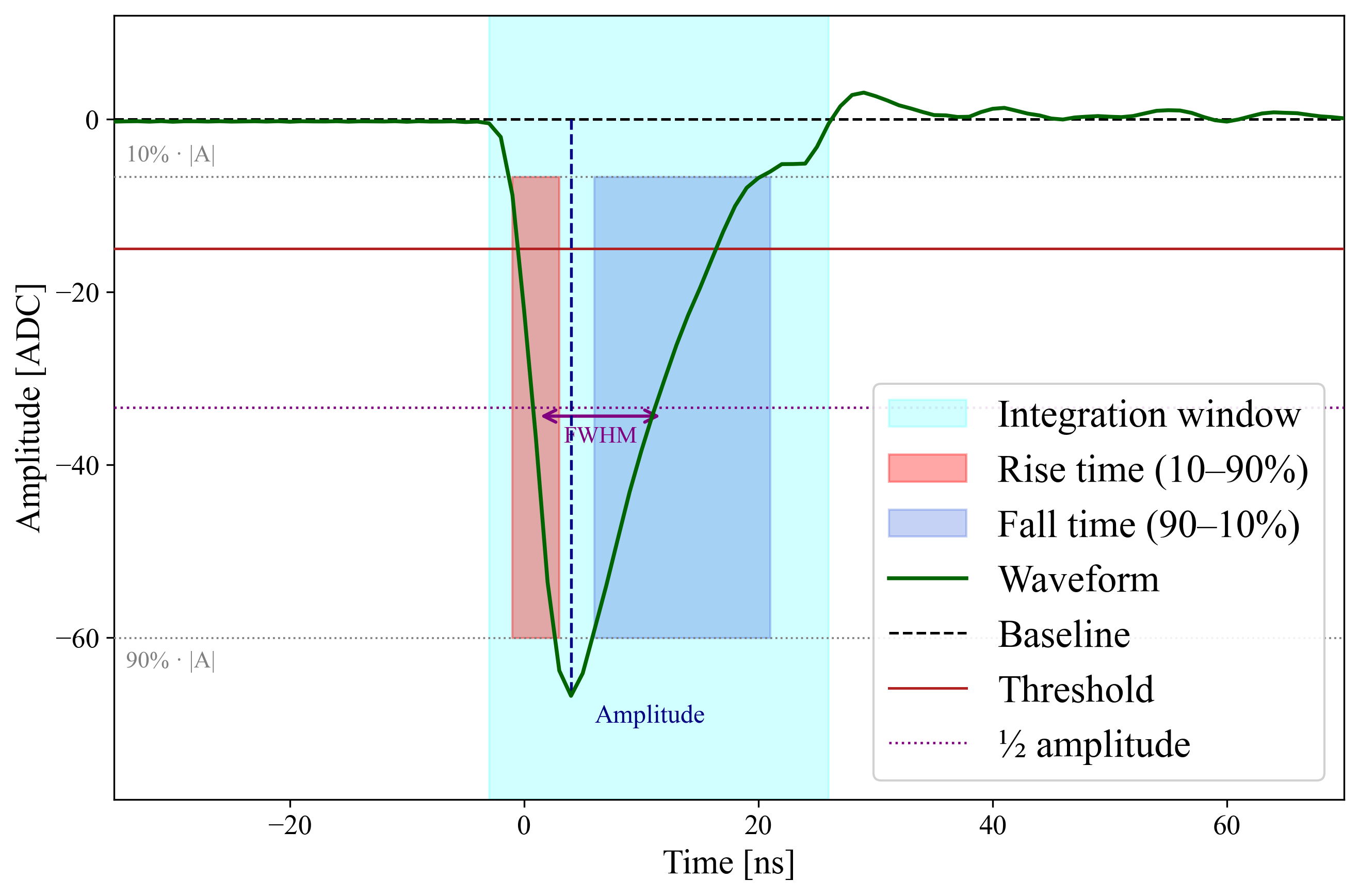}
    \caption{}
\end{subfigure}
\caption{(a) Average waveforms for MCP and dynode PMTs (206,531 waveforms for MCP PMTs, 248,983 waveforms for dynode PMT), with all waveforms aligned to the trigger time (time when the waveform exceeds the 15~ADC counts threshold); (b) Definition of the waveform parameters, including rise time, fall time, width, and the integration window for charge spectrum calculation.}
\label{fig:17}
\end{figure}

To evaluate the PMT signal shape, Figure~\ref{fig:17}(a) presents the measured original dark noise waveforms for MCP and dynode PMTs, averaged from over 200,000 individual waveforms collected in the 6th test. The results show that the average waveform amplitudes are approximately 67~ADC counts (8.9~mV) and 55~ADC counts (7.2~mV) for MCP PMTs and dynode PMTs, respectively, with a threshold---15~ADC counts (2~mV) applied for the individual waveform measurements. Since these waveforms are derived from dark noise, they inherently correspond to single photoelectron (SPE) signals, with MCP PMTs exhibiting larger amplitudes due to their higher gains.

In addition, as shown in Figure~\ref{fig:17}(b), the waveform rise time, fall time, width (FWHM), and charge are defined and calculated. Table~\ref{tab:3} presents these time-related parameters measured in the installation test, along with the Pan-Asia test results for comparison. It can be seen that the values from the two tests are consistent, which further verifies that the PMTs functioned normally after installation.

\vspace{-0.8cm}
\begin{table}[ht]
    \centering
    \caption{PMT waveform parameters}
    \label{tab:3}
    \begin{tabular}{lcccc}
    \toprule
    \multirow{2}{*}{Parameter} & \multicolumn{2}{c}{Installation Test} & \multicolumn{2}{c}{Pan-Asia Test} \\
    \cmidrule(lr){2-3} \cmidrule(lr){4-5}
    & MCP PMT & Dynode PMT & MCP PMT & Dynode PMT \\
    \midrule
    rise-time (ns) & 4.0 & 6.6 & 3.6 & 6.2 \\
    fall-time (ns) & 13.9 & 9.4 & 15.3 & 9.6 \\
    FWHM (ns) & 11.1 & 11.8 & 9.9 & 9.6 \\
    \bottomrule
    \end{tabular}
\end{table}

\begin{figure}[H]
\centering
\begin{subfigure}[b]{0.48\textwidth}
    \includegraphics[width=\textwidth]{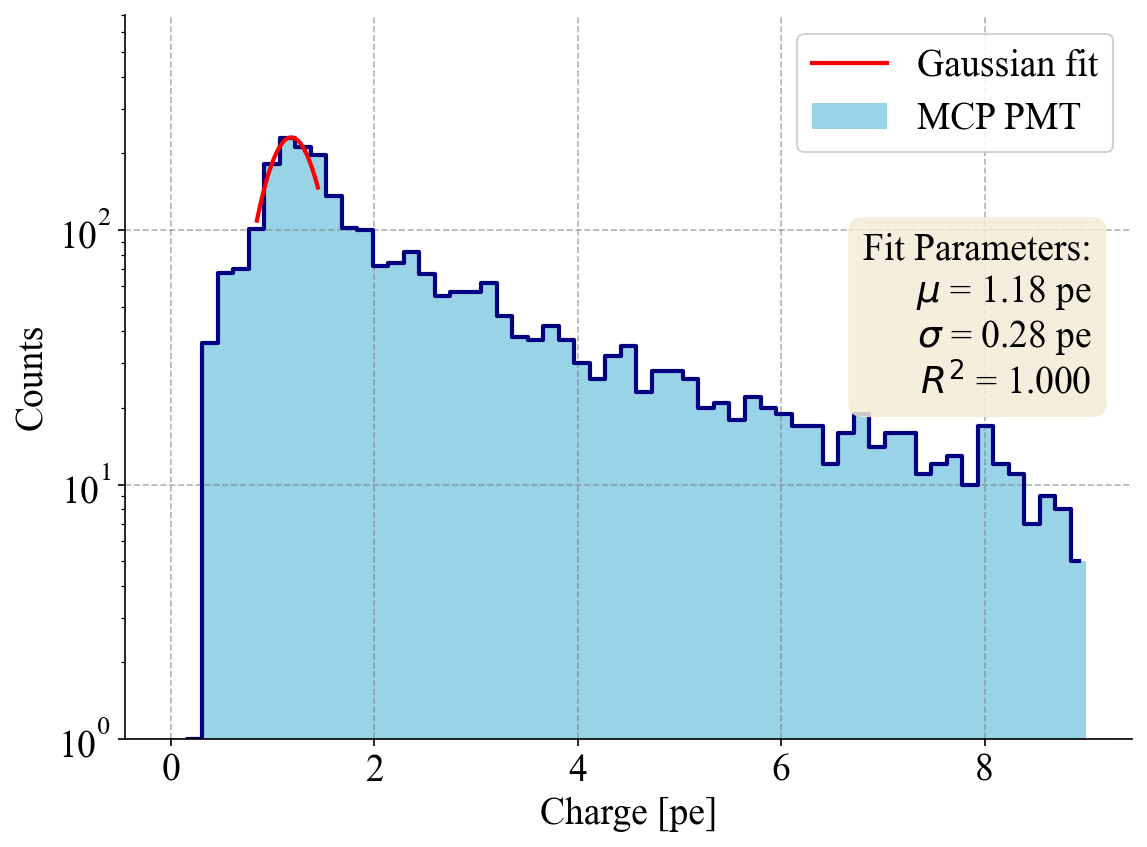}
    \caption{}
\end{subfigure}
\hfill
\begin{subfigure}[b]{0.48\textwidth}
    \includegraphics[width=\textwidth]{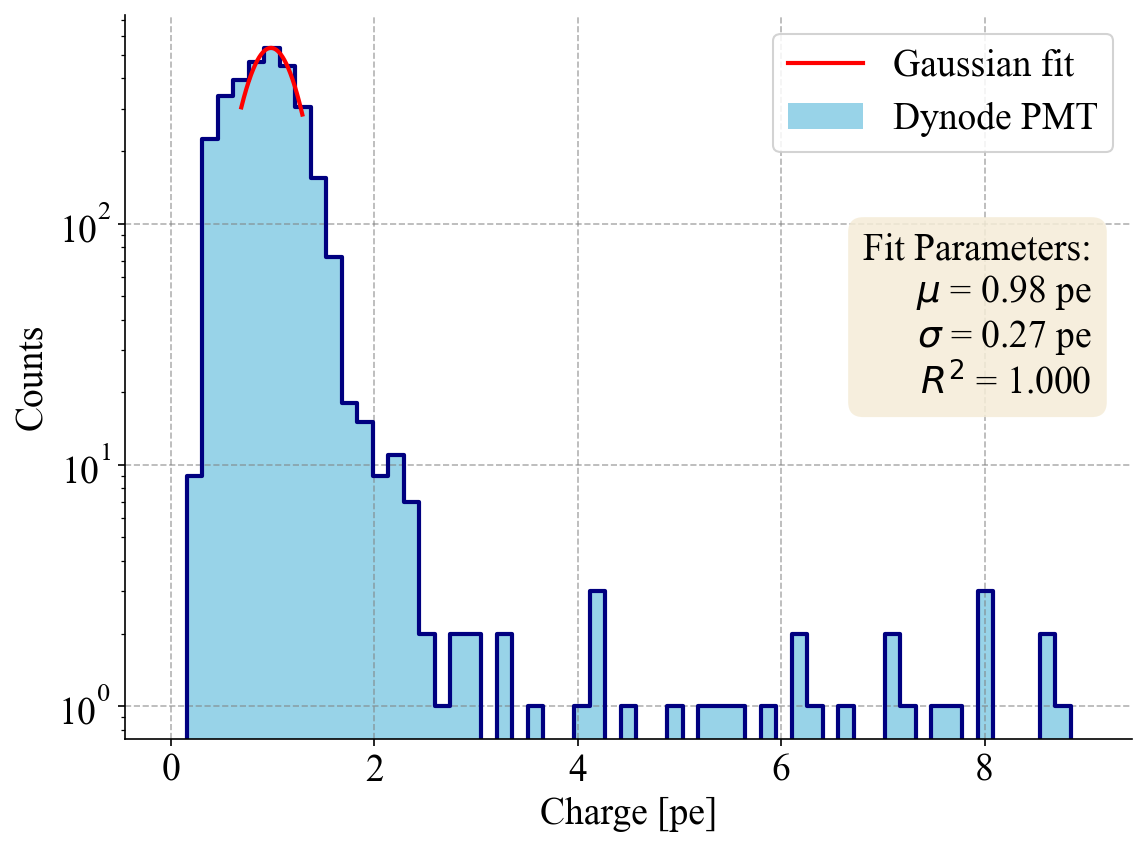}
    \caption{}
\end{subfigure}
\caption{Typical PMT charge spectra, with the Gaussian fitting range defined as $\pm$0.5~pe around the max-count bin. (a) Charge spectrum of an example MCP PMT; (b) Charge spectrum of an example Dynode PMT.}
\label{fig:18}
\end{figure}

Furthermore, the PMT charge spectrum was examined by integrating the collected individual dark noise waveforms, as shown in Figure~\ref{fig:18}. A simple Gaussian fit was performed on these spectra to extract the peak value, with the charge unit converted from ADC$\times$ns to pe (photoelectron, 1~pe $\approx$ 666.67~ADC$\times$ns), yielding approximately 1.18~pe for MCP PMTs and approximately 0.98~pe for dynode PMTs. The deviation from the 1~pe expectation can be attributed to the same factors affecting the PMT gain, as already discussed in Section~\ref{sec:gain}. In addition, a typical long tail in the charge spectrum of MCP PMTs can be observed, which is a well-known feature of this PMT type and has been extensively addressed in the literature~\cite{WENG2024169626,MCP_longtail,MCP_longtail_LSP}.

\section{Conclusions}

This paper presents the test of JUNO 20-inch PMTs during their installation. Seven test campaigns covering the vast majority of the PMTs (19843 out of 20343) were completed within about two years. A brief summary of the PMT installation is also provided as context for the test. Due to the tight installation schedule, these tests were conducted at night with limited testing time. The test protocol and setup are described, with emphasis on light blocking, high-voltage application strategy, and data readout. For each test, prompt results based on 3D PMT maps are presented, together with subsequent analyses of the basic PMT performance parameters including gain, DCR, waveform and charge spectra.

The prompt results verified that the PMTs worked properly. Although the DCR values were relatively high, they were acceptable at this stage due to imperfect light blocking and short cool-down time. The results also indicated no major problems from installation, with only four PMTs found to have no response and two of them successfully recovered. The subsequent performance analysis further confirmed that the PMTs performed as expected and the results were understandable: the waveform shape is consistent with the earlier Pan-Asia test; the deviations of gain and charge-spectrum peak from the expected values are attributed to the change of test conditions relative to previous test; and the long tail observed in the charge spectrum of MCP PMTs is a well-known characteristic of this PMT type. In addition, the experience gained from these tests laid a solid foundation for subsequent detector commissioning and official operation. Overall, these tests played an important role in the successful installation of the PMTs and in ensuring the excellent performance of the final JUNO detector.

\section*{Acknowledgements}

The authors would like to acknowledge the readout electronics, DAQ, DCS, and commissioning team for their close collaboration and support throughout the PMT installation and testing.

This work is supported by the Strategic Priority Research Program of the Chinese Academy of Sciences (Grant No. XDA10000000).

\section*{Declarations}

\textbf{Conflict of interest} On behalf of all authors, the corresponding author states that there is no conflict of interest.
\bibliography{reference}

\end{document}